\documentclass[osajnl,twocolumn,showpacs,superscriptaddress,10pt,floatfix]{revtex4-1} 
\usepackage{amsmath,amssymb,graphicx}
\begin{document}

\title{Dispersion engineered silicon nitride waveguides by geometrical and refractive-index optimization}

\author{J.M. Chavez Boggio}\email{Corresponding author: jboggio@aip.de}
\author{D. Bodenm\"uller}
\author{T. Fremberg}
\author{R. Haynes}
\author{M.M. Roth}
\affiliation{innoFSPEC-VKS, Leibniz-Institut f\"ur Astrophysik Potsdam (AIP), An der Sternwarte 16, D-14482 Potsdam, Germany}

\author{R. Eisermann}
\author{M. Lisker}
\author{L. Zimmermann}
\affiliation{IHP, Im Technologiepark 25, 15236 Frankfurt (Oder), Germany}

\author{M. B\"ohm}
\affiliation{innoFSPEC-InFaSe, University of Potsdam, Am M\"uhlenberg 3, D-14476 Golm, Germany}

\begin{abstract}Dispersion engineering in silicon nitride ($Si_X N_Y$) waveguides is investigated through the optimization of the 
waveguide transversal dimensions and refractive indices in a multi-cladding arrangement. Ultra-flat dispersion of -84.0 $\pm$ 0.5 ps/nm/km 
between 1700 and 2440 nm and 1.5 $\pm$ 3 ps/nm/km between 1670 and 2500 nm is numerically demonstrated. It is shown that typical 
refractive index fluctuations as well as dimension fluctuations during the fabrication of the $Si_X N_Y$ waveguides are a limitation for obtaining 
ultra-flat dispersion profiles. Single- and multi-cladding waveguides are fabricated and their dispersion profiles measured (over 
nearly 1000 nm) using a low-coherence frequency domain interferometric technique. By appropriate thickness optimization, the 
zero-dispersion wavelength is tuned over a large spectral range in both single-cladding waveguides and multi-cladding waveguides 
with small refractive index contrast (3 \%). A flat dispersion profile with $\pm$ 3.2 ps/nm/km variation over 500 nm is obtained 
in a multi-cladding waveguide fabricated with a refractive index contrast of 37 \%. Finally, we generate a nearly three-octave supercontinuum 
in this dispersion flattened multi-cladding $Si_X N_Y$ waveguide.
\end{abstract}

\maketitle 

\section{Introduction}

The development of photonic devices using silicon as the optical medium is paving the way for a monolithically integrated 
optoelectronic platform in a single chip \cite{1,2,3}. Owing to the large nonlinear refractive index of silicon (500 times 
larger than silica), a number of functionalities based on nonlinear interactions have been demonstrated in recent years, such as 
Raman lasers, frequency comb generators, parametric amplifiers, miniature oscilloscopes, wavelength converters, 
etc \cite{4,5,6,7,8,9,10,11,12,13,14,15}. However, silicon is affected by two-photon absorption (TPA) and free carrier 
absorption (FCA), two undesirable phenomena that limit the amount of power that can propagate in the silicon waveguide, so that 
TPA and FCA consequently impair the performance of such a material for nonlinear devices \cite{5}. One way to solve this problem 
is to employ other materials and/or engineer the chromatic dispersion of the waveguide to substantially reduce the power threshold 
for nonlinear interactions.

In most nonlinear applications (supercontinuum generation, parametric interactions, etc) the most suitable dispersion is low and 
anomalous over a broadband spectral range, however other applications like frequency comb generation require moderate anomalous 
dispersion in order to minimze noise buildup \cite{15}. Furthermore, on chip nonlinear pulse compression can require strong normal 
dispersion over a broadband spectral range. Indeed, chromatic dispersion can be modified by changing the core transversal dimensions 
of the waveguide \cite{16,17}. However, this procedure has proven to be insufficient if the aim is to control the dispersion profile 
in an arbitrary way. It has been proposed \cite{18} that adding a thin extra layer of silicon nitride on top of the silicon core 
can modify the dispersion of the silicon waveguide. This approach has been explored further by using different refractive index 
arrangements in a multi-cladding configuration \cite{19,20,21,22,23,24,25,26,27}. The first experimental demonstration of dispersion 
engineering using more than one cladding layer was recently achieved by adding a 60-nm thick Hafnium dioxide layer on top of a silicon 
nitride core, the outcome being that the dispersion became more anomalous \cite{28}. Although there have been considerable advancements in the field of 
silicon photonics over the last years \cite{19,20,21,22,23,24,25,26,27}, a systematic investigation of how to engineer chromatic dispersion is 
still not very advanced in comparison to, for example, the engineering of the dispersion achieved in photonic crystal fibers \cite{29,30,31,32}. 
Furthermore, due to the high index contrast, silicon waveguides have a strong dispersion (larger than hundreds of ps/nm/km) so that 
their engineering is strongly sensitive to refractive index and dimension inaccuracies during the fabrication process. Silicon nitride 
constitutes a promising alternative material to silicon due to the absence of TPA and FCA at all usable infrared wavelengths \cite{23,33,34} and 
its compatibility with the cost-effective CMOS technology. Due to the smaller refractive index contrast with silica, $Si_X N_Y$ appears 
to be a more appropriate material for dispersion engineering than silicon, since typical fabrication inaccuracies will have lower impact on 
dispersion, as we demonstrate in the following. The aim of this investigation is producing flat dispersion profiles over broad spectral ranges 
with arbitrary average dispersion (changing continuously from normal to anomalous dispersion) and with waveguide parameters that are realistic from a
fabrication point of view.

\section{Dispersion engineering approach}

Figure 1 shows the schematics of the cross-section of the waveguides under investigation. There is a substrate layer of 
silicon (not shown in Fig. 1) with a silica buffer layer of two micron thickness on top of it. These are topped with a silicon 
nitride core (with width $w_c$ and height $h_c$) and a silica upper cladding. Figure 1(a) shows the conventional single-cladding silicon 
nitride waveguide without any extra cladding layers. In order to engineer the dispersion, part of the upper cladding is modified 
by including two thin (up to 400 nm) cladding layers, made of silica and/or $Si_X N_Y$. The thicknesses of these extra layers 
are $t_1$ and $t_2$, and their refractive indices are $n_1$ and $n_2$. The lateral wall thickness is $0.3 \times\ t_i$ in all the cases.
Two different refractive index arrangements 
involving the two extra cladding layers are investigated. In the first case, as shown in Fig. 1(b), a high refractive index contrast 
is considered: for this purpose, the first cladding layer is made of silica and covers only the top of the core, while the second layer 
is made of $Si_X N_Y$. In the second case, as shown in Fig. 1(c), a small refractive index contrast is arranged: the two cladding layers are either 
$SiO_2$ or $Si_X N_Y$, however there is a refractive index difference of up to 3\% between each layer (see Fig. 1c). For these two 
different refractive index arrangements, four parameters are changed to optimize the dispersion profile: $w_c$, $h_c$, $t_1$, 
and $t_2$. Additionally, we analyze whether the inclusion of a rib-like waveguide structure, as schematically shown in Figs. 1(d,e,f), 
has further effects on the dispersion profile. We finally investigate the effect of the core side-wall angle, which 
we varied between $78^\circ$ and $90^\circ$.

\begin{figure}[htbp]
\centerline{\includegraphics[width=1.0\columnwidth]{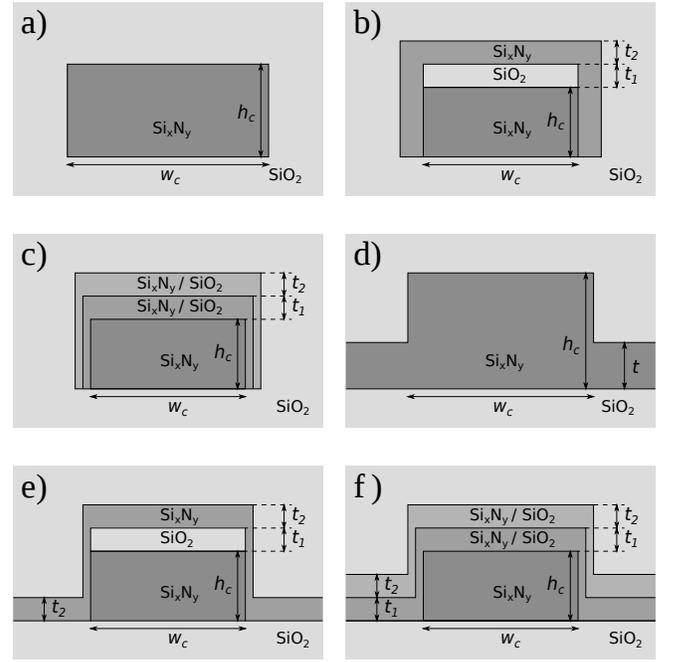}}
\caption{Schematic of the waveguide cross-sections for dispersion engineering investigated in this paper.}
\label{fig:Figure1}
\end{figure}

For the modeling, we employed full vectorial Beam PROP and FemSim packages from commercially available RSoft software. In the first step, we 
checked the consistency of the calculated dispersion profiles when using both packages, obtaining very good agreement between them \cite{35}.
Then, we checked the accuracy of the simulation tool by comparing it with well-known full vectorial analytical solutions 
of chromatic dispersion in cylindrical optical fibers with a $Si_X N_Y$ core. Since the numerical accuracy of the algorithm depends on the 
propagation length and the grid resolution, the simulation was repeated while one of those parameters was increased. Convergence is reached 
when the effective refractive index stabilizes into a particular value.

The effective refractive index is calculated for the quasi-TE and quasi-TM modes of the designed $Si_X N_Y$ waveguides. The chromatic 
dispersion, $D$, is related to the effective refractive index by 
\begin{equation}
D=\frac{-\lambda}{c}\frac{d^2 n_\text{eff}}{d \lambda^2}
\end{equation}
where $\lambda$ is the wavelength, $c$ is the speed of light, and $n_\text{eff}$ is the effective refractive index. The contribution of 
material dispersion to the total dispersion is calculated by using the $Si_X N_Y$ refractive index in equation (1).

\begin{figure}[tbp]
\centerline{\includegraphics[width=1.0\columnwidth]{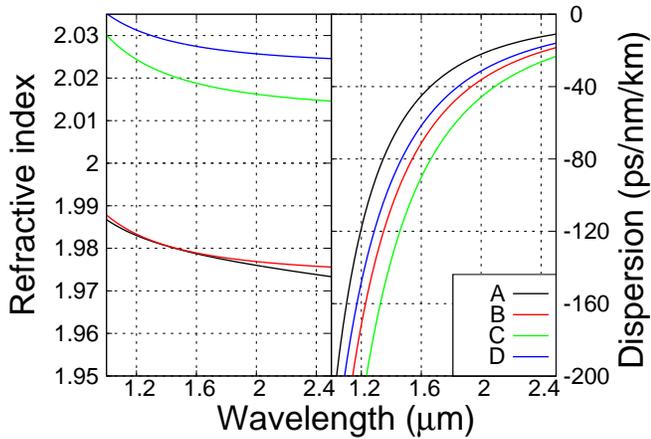}}
\caption{Left: Examples of four silicon nitride refractive index profiles A, B, C, and D. Right: Calculated material dispersion from refractive indices A, B, C and D.}
\label{fig:Figure2}
\end{figure}

An interesting characteristic of silicon nitride is that its refractive index value can be varied over a fairly large range (1.7 - 2.2)
by changing the deposition process. To estimate the impact that refractive index variability has on material dispersion, 
we compared four different values for the silicon nitride refractive index, as shown in Fig. 2(left). The data depicted with a black 
line was retrieved from reference \cite{12} and will be called index A, the red line (index B) comes from the Sellmeier equation 
reported in \cite{36}, the green line (index C) was measured using inline ellipsometry in our fabricated plasma enhanced chemical vapor 
deposition (PECVD) $Si_X N_Y$ films, and finally the blue line (index D) comes from \cite{37}. Figure 2(right) shows the calculated material 
dispersion for the $Si_X N_Y$ refractive indices A, B, C, and D. At short wavelengths, the material dispersion is strongly
dependent on the $Si_X N_Y$ refractive index characteristics (at 1400 nm, for example, the material dispersion is -75 ps/nm/km for index A 
and -135 ps/nm/km for index C). This finding indicates that, unlike in the case of silica or silicon, the dispersion can also be engineered 
by adjusting the deposition process. Due to this variability, the exact refractive index needs to be known in advance with considerable 
accuracy in order to obtain any specific dispersion profile. In what follows of this paper, unless indicated otherwise, all calculations 
are performed using the refractive index C and for the fundamental mode. The goal of our calculations is finding the flattest profiles 
over the largest bandwidth, where the flatness is defined as the smallest variation of the dispersion over a given bandwidth. 

\section{Dispersion engineering in single-clad waveguides by simple geometrical optimization}

We first review the well known case when no extra cladding layers are added ($t_i = 0$ for $i = 1,2$) and the dispersion is 
modified solely by changing the width and the height of the $Si_X N_Y$ core \cite{12}. Since fabricating large $h_c$ values 
can result in an increase of stress, we only consider the case when $w_c$ is larger than $h_c$.The chromatic dispersions calculated for a waveguide 
with width $w_c = 1.8 $ $\mu $m and heights $h_c = 0.75 $ (black), $0.8 $ (red), and $0.85 $ $\mu $m (green) are shown in Fig. 3, 
while Fig. 4 illustrates the dispersion for a waveguide with height $h_c = 0.8 $ $\mu $m and widths $w_c = 1.4$ (black line), $1.8$ (red), and 
$2.2 $ $\mu $m (green). For comparison purposes, the calculations were done for the quasi-TE (solid lines) and quasi-TM (dotted lines) modes.
While the material dispersion is always normal in the near infrared as shown in Fig. 2(right), the strong geometrical dispersion due to the 
confinement of the light in a waveguide results in a total dispersion that can be anomalous over quite large spectral bandwidths \cite{12}. The 
dispersion profiles for quasi-TE modes are flatter and tend to shift to longer wavelengths in comparison to the quasi-TM mode profiles. 
By increasing the core width, the dispersion is vertically down shifted and becomes flatter - an effect that is much stronger 
for the quasi-TE mode. Furthermore, a small increase of the core height produces a fairly large vertical up-shift of 
the dispersion profile, while its shape remains unchanged. From the results in Figs. 3 and 4, a basic rule for dispersion engineering 
can be deduced: a dispersion profile with a given flatness can be vertically shifted by changing $h_c$.

\begin{figure}[htbp]
\centerline{\includegraphics[width=1.0\columnwidth]{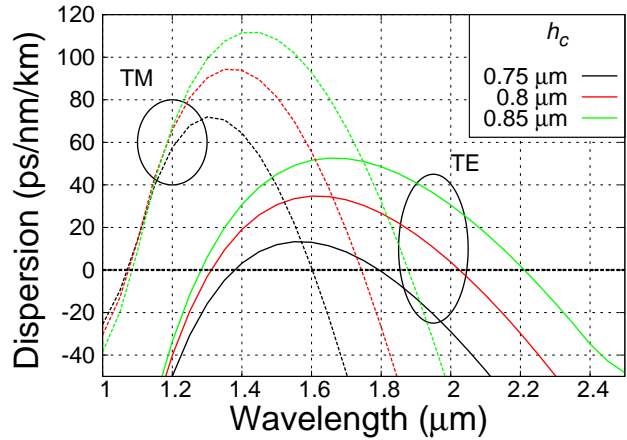}}
\caption{Dispersion vs. wavelength for $w_c = 1.8 $ $\mu $m and $h_c = 0.75$, $0.8$, and $0.85 $ $\mu $m. Quasi-TE (TM) mode: solid (dashed) lines.}
\label{fig:Figure3}
\end{figure}

\begin{figure}[tbp]
\centerline{\includegraphics[width=1.0\columnwidth]{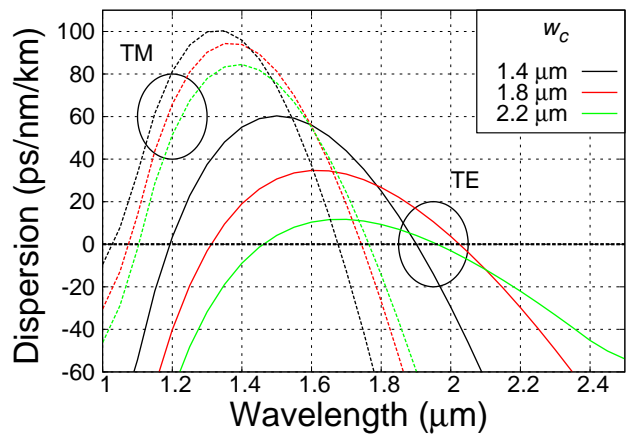}}
\caption{Dispersion vs. wavelength for $h_c = 0.8 $ $\mu $m and $w_c = 1.4$, $1.8$, and $2.2 $ $\mu $m. Quasi-TE (TM) mode: solid (dashed) lines.}
\label{fig:Figure4}
\end{figure}

In order to analyze the number of propagating modes supported by the waveguide with  $h_c = 0.8 $ and $w_c = 1.8$ $\mu $m, the effective refractive indices for the fundamental (1$st$) and 
high-order modes are calculated for both the quasi-TE (Fig. 5-left) and the quasi-TM (Fig. 5-right) cases. A mode is supported if its effective
refractive index value lies between the silicon nitride (red solid line) and silica (black solid line) refractive index values. Note that at 1550 nm,
four modes are supported respectively for both the quasi-TE case and the quasi-TM case. The waveguide is single mode at wavelengths beyond $2.5 $ $\mu $m, 
while at $1 $ $\mu $m, up to ten modes can propagate. In what follows of this paper, unless indicated otherwise, the calculations refer to the quasi-TE mode.

\begin{figure}[tbp]
\centerline{\includegraphics[width=1.0\columnwidth]{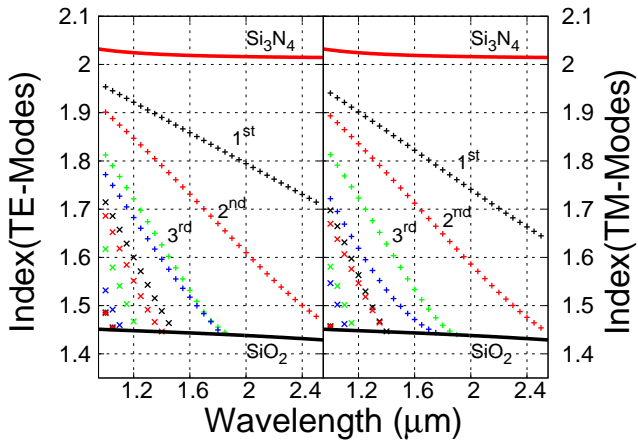}}
\caption{Effective refractive index for the quasi-TE (Left) and quasi-TM (Right) supported modes. The refractive index of $Si_X N_Y$ ($Si O_2$) is in a red (black) solid line.}
\label{fig:Figure5}
\end{figure}
 
\section{Dispersion engineering in multi-cladding waveguides with a large ($\sim 1.4$) refractive index contrast}
To further engineer the chromatic dispersion, we investigate the arrangement shown in Fig. 1(b), where the cladding layers have a
refractive index contrast of $\sim 1.4$: the first cladding is made of silica whereas the second is a silicon nitride layer, while everything is covered 
by a silica upper cladding. In order to avoid unrealistic dimension values that could produce extremely flat dispersion profiles but 
could hardly be fabricated, we changed $w_c$, $t_1$, and $t_2$ in steps of 10 nm while $h_c$ is changed in steps of 25 nm. As illustrated in 
Fig. 6, the optimization of the four parameters produces very flat profiles across the 1.65 - 2.5 $\mu $m spectral region with average dispersion 
that can be tuned from -85 to +17 ps/nm/km. The third order dispersion is calculated to be $S_D$ = 0.0005 ps/nm$^2$/km at 2 microns 
for most of the dispersion cases. The dispersion flatness in Fig. 6 is comparable with the flattest dispersion profiles found in photonics 
crystal fibers \cite{29,30,31,32}.

Table 1 shows the optimized parameters ($h_c$, $w_c$, $t_1$, and $t_2$) producing the flattened profiles shown 
in Fig 6. The one with average dispersion of -85 ps/nm/km is called profile $A$ while the one with +17 ps/nm/km is named profile $K$. The remarkable 
fact from Table 1 is that for a given $h_c$, a flattened profile can always be engineered by optimizing the $w_c$, $t_1$, and $t_2$ values. The 
optimized values of $w_c$, $t_1$, and $t_2$ are 'universally' valid in the sense that an increase of the value of $h_c$ invariably produces a 
vertical up-shift of the dispersion, which nonetheless leaves the flatness of the dispersion profile nearly unchanged. Once the optimized $h_c$, 
$w_c$, $t_1$, and $t_2$ values for flat dispersion are found, a change of $h_c$ is compensated by small changes in the other parameters. By 
repeating the simulations with other refractive indices, it is found that a vertical up-shift (down-shift) of the profile is always due to 
an increase (decrease) of $h_c$. However, as can be expected, the optimal parameters that produce flat dispersion profiles change as the 
refractive index is changed. 

\begin{figure}[tbp]
\centerline{\includegraphics[width=1.0\columnwidth]{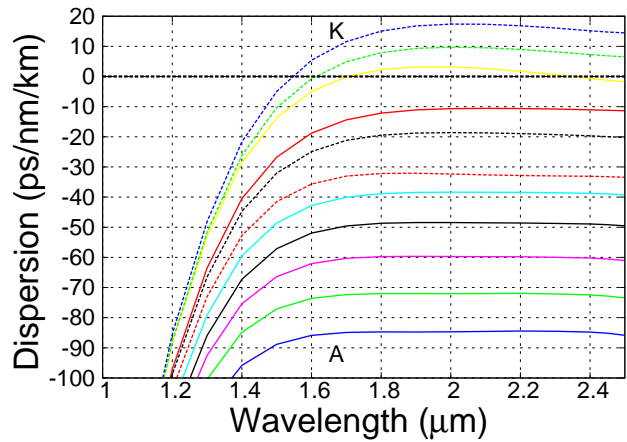}}
\caption{Dispersion flattened profiles after optimizing $h_c$, $w_c$, $t_1$, and $t_2$. Increasing $h_c$ changes the profile from A to K with minor adjustements of $w_c$, $t_1$, and $t_2$.}
\label{fig:Figure6}
\end{figure}

\begin{table}[h!]
  \caption{Optimized parameters for flat dispersion profiles.}
  \begin{center}
    \begin{tabular}{cccccccccccc}
    \hline
     & A & B & C & D & E & F & G & H & I & J & K\\
    \hline
    $w_c$ & 1.65 & 1.7 & 1.7 & 1.75 & 1.7 & 1.7 & 1.65 & 1.7 & 1.65 & 1.7 & 1.7\\
    $h_c$ & 0.775 & 0.75 & 0.8 & 0.825 & 0.85 & 0.875 & 0.9 & 0.925 & 0.95 & 0.975 & 1.0\\
    $t_1$ & 0.25 & 0.25 & 0.25 & 0.25 & 0.25 & 0.26 & 0.25 & 0.25 & 0.25 & 0.25 & 0.25\\
    $t_2$ & 0.21 & 0.2 & 0.21 & 0.21 & 0.22 & 0.23 & 0.23 & 0.23 & 0.22 & 0.22 & 0.22\\
    \hline
    \end{tabular}
  \end{center}
\end{table}

The black line in Figure 7 shows the calculated effective mode-area of a multi-cladding waveguide with identical parameters than the one labelled 
as C in Fig. 6. For comparison, the red line illustrates the different result obtained when substituting the silica cladding layer by one of 
silicon nitride. On the right side of Fig. 7, the mode profiles at a wavelength of $1.55 $ $\mu $m are depicted together with the waveguide 
structure, showing that the field is tightly confined in the core whenever the silica extra cladding layer is included.

\begin{figure}[tbp]
\centerline{\includegraphics[width=1.0\columnwidth]{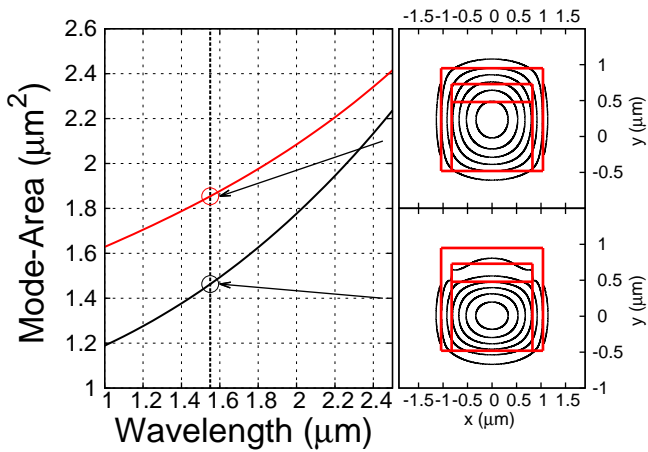}}
\caption{Left: Effective mode area for the profile C in Fig. 6 (black line) and when the silica cladding layer is substituted by $Si_X N_Y$ (red line). Right: Mode profiles.}
\label{fig:Figure7}
\end{figure}

\begin{figure}[htbp]
\centerline{\includegraphics[width=1.0\columnwidth]{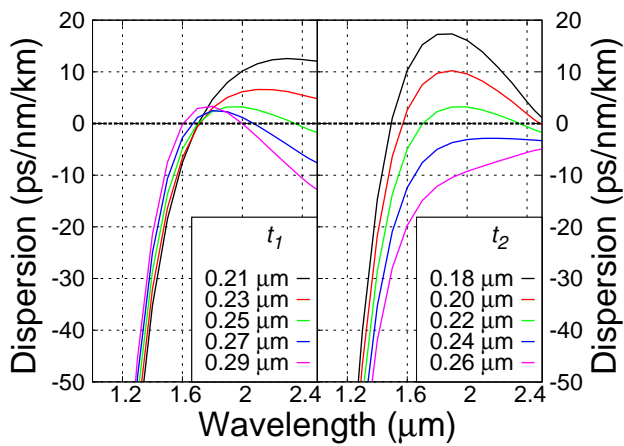}}
\caption{Chromatic dispersion as $t_1$ is changed from $0.21$ to $0.29 $ $ \mu $m (Left) and $t_2$ from $0.18$ to $0.26 $ $ \mu $m (Right).}
\label{fig:Figure8}
\end{figure}

In order to facilitate the design and fabrication of optimized dispersion profiles, it is important to know whether or not it is 
possible to find simple rules that allow for a predictable modification of the chromatic dispersion. We first checked how dispersion is 
modified by changing the thickness of the silica cladding layer from $t_1 = 0.21$ to $0.29 $ $\mu $m, as shown in Fig. 8(left). The 
dispersion profile in green is the same as the one depicted in yellow (profile I) in Fig. 6. As $t_1$ is increased, the dispersion is tilted 
up for shorter wavelengths, and down for longer wavelengths, while all dispersion profiles intersect at a narrow wavelength region 
(around $1.75 $ $\mu $m) \cite{20,25}. We numerically verified that this intersection point can be tuned over a broad spectral range by changing
$h_c$, $w_c$, $t_1$, or $t_2$. On the other hand, by increasing $t_2$ the dispersion is vertically down-shifted (the black line 
corresponds to $t_2 = 0.18 $ $\mu $m, while the magenta one refers to $t_2 = 0.26 $ $\mu $m). Since the down-shift of the profile is less pronounced
at longer wavelengths, the dispersion flatness can be controlled via a tuning of $t_2$ \cite{20,25}. Results in Fig. 8 show that a proper adjustement of 
$t_1$ and $t_2$ produces a flattened dispersion at spectral regions that can be tailored.

The advantage of using $Si_X N_Y$ as the core material instead of silicon is mainly due to its robustness against dimension fluctuations.
Silicon slot waveguides were shown to require fabrication accuracies of the order of 1 nm in order to have a controllable dispersion tuning \cite{20,24,25}. 

\section{Dispersion engineering in multi-cladding waveguides with small refractive index contrast}

The flat dispersion profiles found in the previous section were obtained using silica and silicon nitride as first and 
second cladding layers, giving a ratio of their refractive indices of $n_2/n_1 = 1.37$. An important question in this
context is how this ratio affects the performance of the engineering process \cite{24}. Therefore, a low-contrast ($n_2/n_1 = 0.985$) 
cladding-refractive-index arrangement is investigated: for the first cladding layer $n_1 = 0.985\times n_{Si_X N_Y}$ while for 
the second cladding layer $n_2 = 0.97\times n_{Si_X N_Y}$, with $n_{Si_X N_Y}$ being the silicon nitride refractive index of 
the core. The thicknesses $t_1$ and $t_2$ are varied from $0.1$ to $0.4 $ $\mu $m, and the values of $h_c$ and $w_c$ are scanned 
to retrieve flat dispersion profiles. Fairly flat profiles were found only for a reduced range of parameters. Figure 9 shows 
in solid lines those dispersion profiles plotted for $t_1 = t_2 = 0.4 $ $\mu $m and $w_c = 2.2 $ $\mu $m, while $h_c$ is varied from $0.5$ 
to $ 0.65 $ $\mu $m. The dispersion is fairly flat from $2.2$ to $2.4 $ $\mu$m, although not to the same extent as it is in the higher 
index contrast case and is not over such a broad bandwidth. In order to understand the real impact of this refractive index 
gradient on dispersion, we perform the calculation for the same values of $h_c$, $w_c$, $t_1$, and $t_2$ while the refractive 
index contrast is reduced to zero (i.e. like a single-cladding waveguide with larger $h_c$ and $w_c$). The simulation results 
are shown in dotted lines for comparison: the two extra cladding layers with refractive indices that are 1.5 and 3 \% lower 
than the core produce a small vertical down-shift of the dispersion profile with only a negligible change of its shape. By
performing simulations where the $w_c$, $t_1$, or $t_2$ parameters are changed, it is found that the retrieved dispersion is 
always a vertically down-shifted version of the case when the refractive index contrast is set to zero: multi-cladding waveguides 
with small refractive index contrast have equivalent dispersion profiles than single-cladding waveguides. This is explained 
as due to the small contribution from geometrical dispersion when the index contrast is small.

The effective mode area as a function of wavelength is plotted in Fig. 10, for both cases (with and without 
refractive index gradient). The field spatial distribution is very similar for both index arrangements, with a noticeable 
difference in the mode-area only at short wavelengths.

\begin{figure}[htbp]
\centerline{\includegraphics[width=1.0\columnwidth]{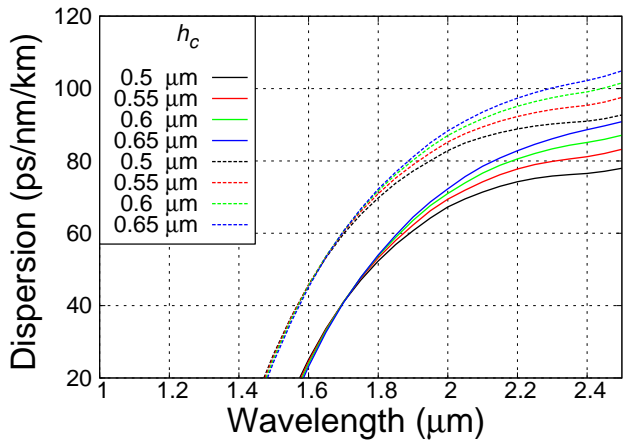}}
\caption{Dispersion as $h_c$ is changed from $0.5$ to $0.65 $ $\mu $m and $t_1 = t_2 = 0.4 $ $\mu $m. Dashed lines show the case when all layers have the same refractive index. }
\label{fig:Figure9}
\end{figure}

\begin{figure}[tbp]
\centerline{\includegraphics[width=1.0\columnwidth]{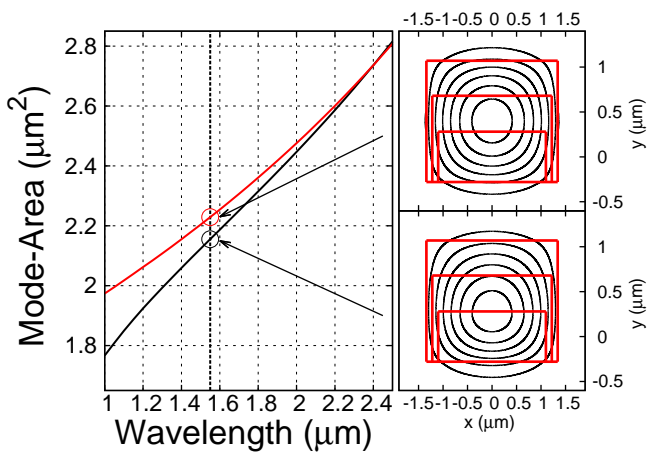}}
\caption{Mode area vs. wavelength: with (black line) and without (red line) refractive index gradient.}
\label{fig:Figure10}
\end{figure}

For completeness, other arrangements involving small refractive index contrasts were analyzed. For example, by choosing 
$n_1 = 0.97 \times n_{SiO_2}$ and $n_2 = 0.985\times n_{SiO_2}$ or $n_1 = 0.97\times n_{Si_X N_Y}$ and $n_2 = 0.985\times n_{SiO_2}$, 
no appreciable improvement is obtained in terms of dispersion control if compared with the single-cladding waveguide \cite{38}. The results
in this section show that moderate to large refractive index contrasts between the different layers of the waveguide
are required in order to enhance the strength of the geometrical dispersion and produce arbitrary dispersion profiles.

\section{Chromatic dispersion of a 'rib-like' waveguide and the effect of the core sidewall angle}
The waveguide geometry can be modified not only by changing $w_c$ and $h_c$, but also by including a rib-like structure as in 
Figures 1(d,e,f). The chromatic dispersion is calculated for a waveguide with a rib-like structure like Fig. 1(f) and having 
$h_c = 0.75 $ $\mu $m, and $t_1 = t_2 = 0.1 $ $\mu $m (dotted lines) and $t_1 = t_2 = 0.4 $ $\mu $m (solid lines). Three cases are considered: 
$w_c = 1.3 $, $w_c = 1.5 $, and $w_c = 1.8 $ $\mu $m. The profiles in dotted lines 
show that the inclusion of the rib-like structure makes the dispersion less sensitive to changes of $w_c$. A comparison of the 
results in solid lines in Fig. 11 with those in solid lines depicted in Fig. 9 shows that the inclusion of a rib-like structure 
produces a vertical down-shift of the dispersion as well as a change of its flattness. Therefore, 
$t$ can be used in conjunction with $h_c$ to vertically shift a dispersion profile that has already 
been engineered. Furthermore, this kind of profile can be used to reduce the impact of dimension fluctuations. On the other hand, we 
verified that the effect of the lateral wall ($0.3 \times\ t_i$) is smaller than the effect due to a rib-like structure.

\begin{figure}[htbp]
\centerline{\includegraphics[width=1.0\columnwidth]{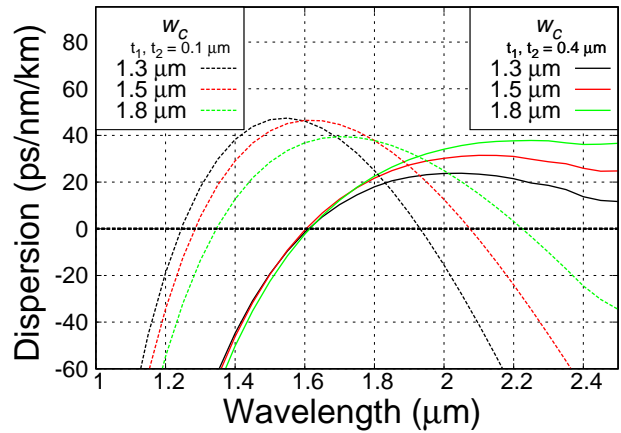}}
\caption{Chromatic dispersion for $t_1 = t_2 = 0.1 $ $\mu $m (dotted line) and $t_1 = t_2 = 0.4 $ $\mu $m (solid line).}
\label{fig:Figure11}
\end{figure}

While our numerical analysis considered perfectly rectangular structures, during the fabrication process the resulting waveguides 
are trapezoidal-like rather than rectangular. As can be seen in the inset of Fig. 12 where the scanning electron microscope (SEM) 
image of a $Si_X N_Y$ waveguide is depicted, in our fabricated waveguides the angle is typically $82^\circ$ rather 
than $90^\circ$. To evaluate the impact of the core side-wall angle, the dispersion is calculated for a single-clad $Si_X N_Y$ 
waveguide with $h_c = 0.8 $ $\mu $m and $w_c = 1.8 $ $\mu $m. As the angle increases, the dispersion profile is vertically 
up-shifted and somewhat less flattened. Even though the change is small, it is not negligible and might introduce uncertainty
when designing engineered profiles.

\begin{figure}[htbp]
\centerline{\includegraphics[width=1.0\columnwidth]{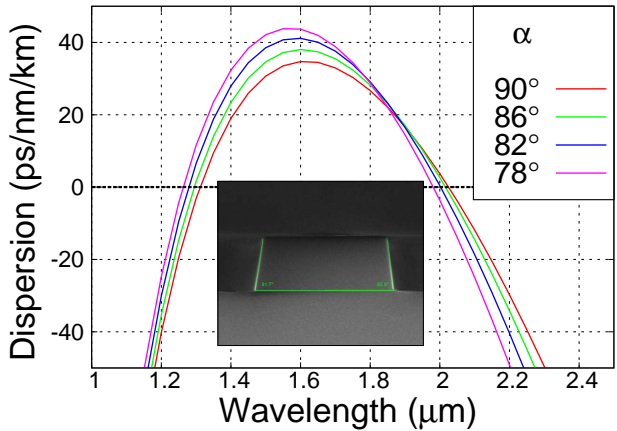}}
\caption{Chromatic dispersion profile as the sidewall angle is changed from $78^\circ$ to $90^\circ$. Inset: SEM image of fabricated waveguide with typical $82^\circ$ sidewall angle.}
\label{fig:Figure12}
\end{figure}

\section{Impact of variations of the refractive index and geometrical dimensions}
\subsection{Fabrication inaccuracies}

Due to unavoidable uncertainties during the fabrication process of the $Si_X N_Y$ waveguides, the refractive index and the 
transversal dimensions might not correspond to what was originally conceived. Consequently, if the fabrication inaccuracies are
too large, the designed ultra-flat dispersion profile might not be as flat as initially intended. To assess the impact that refractive 
index and transversal dimension random fluctuations have on dispersion, several PECVD $Si_X N_Y$ and $Si O_2$ films have been grown 
with a specific thickness and with an uniform refractive index. Both the refractive index and the thickness are measured using 
in-line ellipsometry at locations evenly scattered over a 200 mm diameter wafer. Figure 13 shows the $Si_X N_Y$ refractive index for 
25 wafer locations as well as their corresponding calculated material dispersions. The refractive index fluctuation (around 2\%) 
will provoke a fluctuation (and therefore an uncertainty) of the dispersion of $\sim 20$ ps/nm/km at 1550 nm and $7$ ps/nm/km at 2500 
nm. However, since most of our films exhibited a refractive index fluctuations of 1\% over the entire wafer, this fluctuation shall 
produce a $\sim 10$ ps/nm/km uncertainty at 1550 nm.

Dimensions inaccuracy during fabrication is another (important) source of dispersion uncertainty. While the targeted thickness of the 
$Si_X N_Y$ film was 750 nm, the measured values over the 200 mm wafer scatter from 752 to 776 nm as shown in Fig. 14. However, it is always 
possible to select a region of the wafer where the thickness corresponds to what was originally intended. Nevertheless, the average offset 
of 14 nm is an example of a fairly large uncertainty: grown PECVD $Si_X N_Y$ films with accuracy of 2-3 nm are also very common. These 
values indicate that the dispersion uncertainty caused by the thickness inaccuracies are of the order of a few ps/nm/km. Note also that 
the thickness variation over a waveguide scale of a few cm is just $\sim 2$ nm, therefore no significant dispersion variation might be 
expected along the length of typical waveguides (a few cm).

\begin{figure}[tbp]
\centerline{\includegraphics[width=1.0\columnwidth]{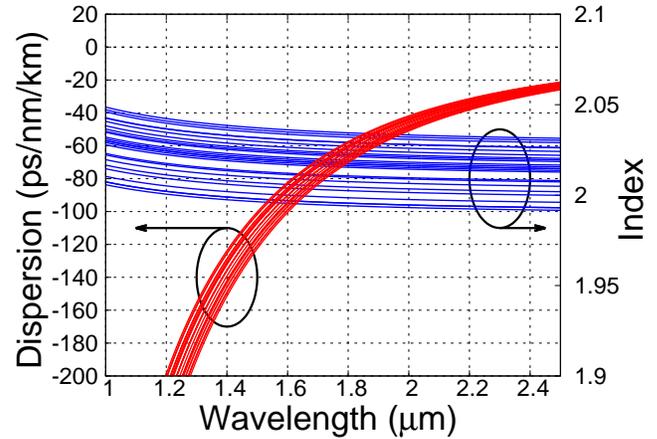}}
\caption{Refractive index and material dispersion calculated at 25 evenly scattered locations over a 200 mm wafer.}
\label{fig:Figure13}
\end{figure}

\begin{figure}[tbp]
\centerline{\includegraphics[width=1.0\columnwidth]{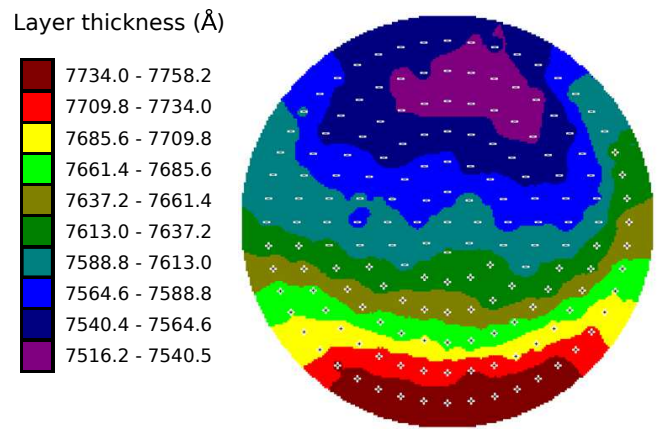}}
\caption{Thickness measured at 169 locations scattered over the 200 mm wafer.}
\label{fig:Figure14}
\end{figure}

\subsection{Chromatic dispersion in multi-cladding waveguides with refractive index fluctuations}

Figure 15 illustrates the impact of refractive index fluctuations on the dispersion flatness in a multi-cladding waveguide with refractive 
index arrangement as in Fig. 1(e). One of the $Si_X N_Y$ refractive index values from Figure 13 is used for the simulations. The flat dispersion profile 
(variation of $\pm$ 1.8 ps/nm/km from 1250 to 2500 nm) illustrated by the black line is obtained with $w_c = 2.1 $ $\mu $m, $h_c = 0.7 $ $\mu $m, 
$t_1 = 0.2 $ $\mu $m, and $t_2 = 0.26 $ $\mu $m. Starting from this flat case, we apply an increase of 3\% of the $Si_X N_Y$ refractive index 
(core and second cladding layer) and a decrease of 3\% of the refractive index in the silica cladding layers. The new dispersion profile 
is depicted with a red line. The flatness is strongly reduced, suggesting that uncontrolled fluctuations of the refractive index during the 
fabrication of the waveguide might drastically 
impair the engineering performance. However, by changing the waveguide parameters to $w_c = 2.1 $ $\mu $m, $h_c = 0.71 $ $\mu $m, 
$t_1 = 0.2 $ $ \mu $m, and $t_2 = 0.24 $ $\mu $m, it is possible to retrieve again a flat profile as depicted by the green line. Since 
a 3\% refractive index variation is a rather extreme case of fluctuation, it can be concluded that while ultra-flat dispersion 
profiles (with variations smaller than $\pm$ 1 ps/nm/km) are unlikely to be produced in practice, flat profiles with $\pm$ 5 ps/nm/km 
variation over hundreds nanometers are realistical to be engineered.

\begin{figure}[tbp]
\centerline{\includegraphics[width=1.0\columnwidth]{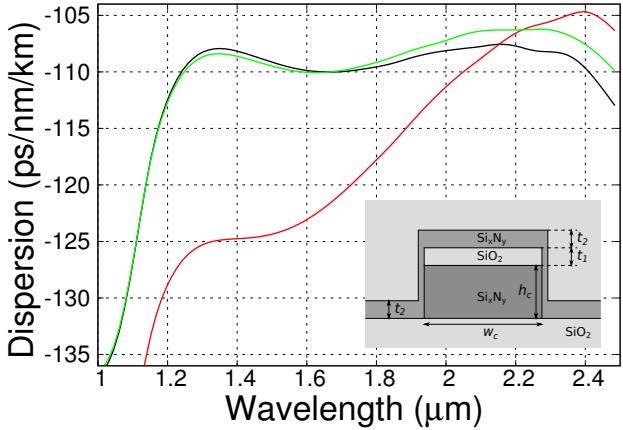}}
\caption{Flat dispersion after optimization of $w_c $, $h_c $, $t_1 $, and $t_2 $ (black line). After an increase/decrease of 3\% of the $Si_X N_Y$/$Si O_2$ refractive index the flatness is lost (red line).
Further optimization reestablish the flatness (green line).}
\label{fig:Figure15}
\end{figure}

Figure 16 illustrates the impact that a random fluctuation of $t_1 $ will have on a flat dispersion profile. The plot in green is the same 
case than the black plot in Figure 15 and corresponds to $t_1 = 0.2 $ $\mu $m. As $t_1$ is changed, the dispersion profile deviates from
the flattest case: a 10 nm deviation of $t_1 $ ps/nm/km reduces the bandwidth of flat dispersion from 1000 nm (green) to 800 nm (magenta) or 620 nm (black). 

\begin{figure}[tbp]
\centerline{\includegraphics[width=1.0\columnwidth]{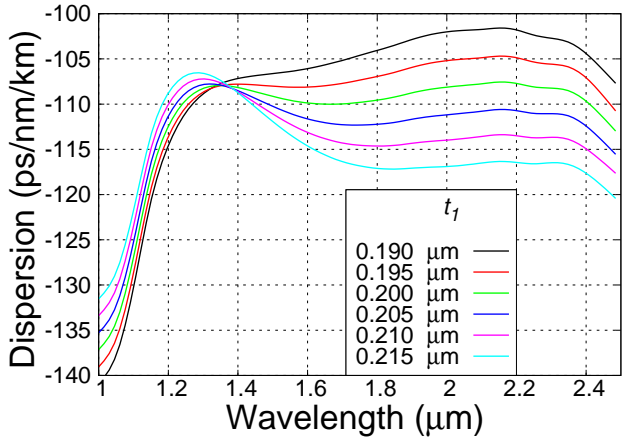}}
\caption{Green line shows the flat dispersion after optimization of $w_c $, $h_c $, $t_2 $, and $t_1 = 0.2 $ $\mu $m. As $t_1 $ is changed from 0.19 to 0.215 $\mu $m the flatness 
is reduced.}
\label{fig:Figure16}
\end{figure}

\section{Dispersion measurements of PECVD silicon nitride waveguides}

\subsection{Experimental procedure}

Single- and multi-cladding $Si_X N_Y$ waveguides having lengths up to 4.85 cm were fabricated using a PECVD process. To 
tune the chromatic dispersion, different core heights (from $h_c = 0.725$ to $h_c = 0.865 $ $ \mu $m), core widths 
($w_c = 1.3$, $1.5$ and $1.8 $ $\mu $m), and thicknesses (from $t_i = 0.04$ to $t_i  = 0.4 $ $\mu $m, $i = 1,2$) were produced. 
The fabricated waveguides have small or large refractive index contrasts. Using a low-coherence frequency domain 
interferometric technique \cite{39}, the chromatic dispersion of these $Si_X N_Y$ waveguides is measured. The experimental 
arrangement is shown in Fig. 16. The light of a supercontinuum source (SC), spanning from $1$ to $2.4 $ $\mu $m, is collimated 
with an objective, while it is polarized with a broadband polarizer. By rotating the polarizer, we excited either the quasi-TE or 
the quasi-TM mode of the waveguide. With a beam splitter, the SC light is splitted into the two arms of the interferometer. 
In one arm (the reference one), a corner cube which is placed on a motorized translation stage allows for the adjustment of 
the optical path length to obtain the interference pattern. The other arm (the sample arm) contains mirrors to redirect the 
SC light into an objective that focuses the light in a lensed fibre, finally coupling it into the waveguide under test. An 
almost identical lensed fibre is placed in the reference arm to compensate for the dispersion of the fibre we use to couple light 
into the waveguide. In this way, we minimize systematic errors in our measurements by avoiding that the low amount of group delay 
due to our $Si_X N_Y$ waveguides is not overwhelmed by the group delay from the lensed fibre. At the waveguide output, 
an objective is used to collect and collimate the light exiting the chip. A linear polarizer is then used to ensure that only 
the TE or TM modes are measured. Finally, the light from both arms is combined using a beam splitter, so that the interference signal 
can be detected using two optical spectrum analyzers (OSAs). The measurement is performed with and without the waveguide in order 
to substract the group delay due to the unbalanced optical components (fibers, beam splitters, objectives, etc.) in the interferometer. 
An infrared camera is used to monitor the Gaussian beam profile of the outcoming light. Even though a few modes can propagate in the waveguide,
most of the power is coupled into the fundamental mode. 

\begin{figure}[htbp]
\centerline{\includegraphics[width=1.0\columnwidth]{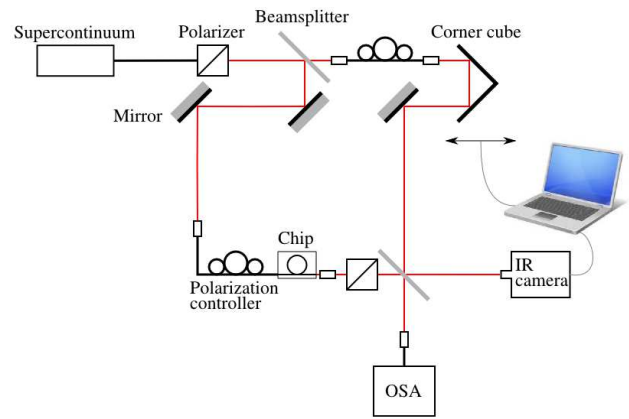}}
\caption{Experimental setup for chromatic dispersion measurements. The red line depicts the free space optical path.}
\label{fig:Figure17}
\end{figure}
\begin{figure}[htbp]
\centerline{\includegraphics[width=1.0\columnwidth]{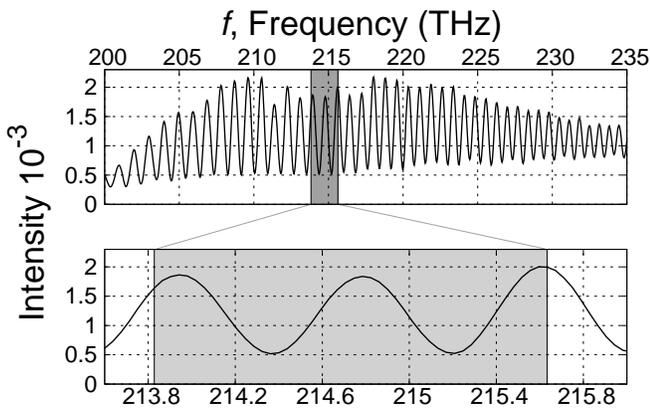}}
\caption{Top: Interference pattern of one of the three spectrally overlapping sections. Bottom: Sub-section where the fitting to obtain $\frac{d \phi}{d f}$ is performed.}
\label{fig:Figure18}
\end{figure}
\begin{figure}[htbp]
\centerline{\includegraphics[width=1.0\columnwidth]{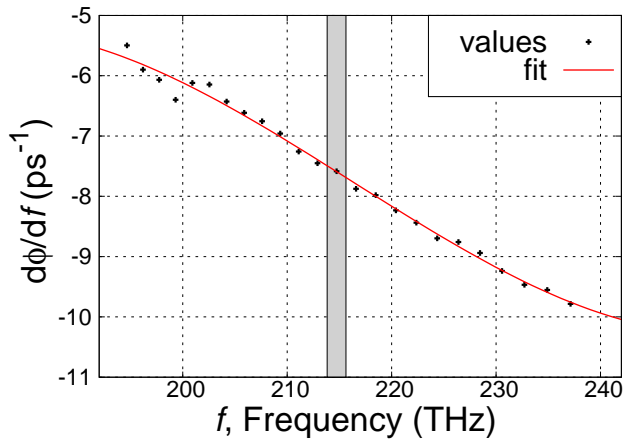}}
\caption{Black dots: measured $d \phi /d f$. Red line: fitting of $d \phi /d f$ using a third order polynomial. Shadow area: sub-section to calculate one of the dots (same as in Fig. 17).}
\label{fig:Figure19}
\end{figure}

An accurate determination of the dispersion profile requires measuring the interference spectrum over the largest bandwidth possible. 
Due to the wavelength dependent response of the optical components, we divided the 1000 nm measurement bandwidth in three overlapping 
spectral sections. Figure 17 shows an example of a typical averaged interference spectrum of one of those three sections. The polarized
SC light is aligned with the quasi-TE propagation mode in a single-cladding $Si_X N_Y$ waveguide (with $L = 4.85 $ cm, 
$h_c = 0.725 $ $\mu $m and $w_c = 1.8 $ $\mu $m). The averaging is done using 100 spectra in order to reduce errors from fluctuations 
of the optical path length in the interferometer. The dependency of the effective refractive index with frequency produces fringes 
which are not equally spaced. The unequally spaced maxima (or minima) account for a phase difference ($\phi$) as a function of 
frequency, which is related to the dispersion through $D = \frac{-2\pi c}{\lambda}\frac{d^2 \phi}{d f^2}$. To calculate the dispersion, 
the spectrum is divided in sub-sections containing a few maxima (three in the case illustrated in Fig. 17), while each subsection is 
fitted with an appropiate function that allows retrieving $\frac{d \phi}{d f}$ values. The black dots in Figure 18 show the 
retrieved $\frac{d \phi}{d f}$ values, while the red curve is the fitting of the $\frac{d \phi}{d f}$ data using a third-order 
polynomial. By taking derivatives of this fitting curve, it is possible to obtain the second- and higher-order dispersion curves for 
the fabricated waveguides. The Appendix describes in more detail the fitting procedure and the estimated error in the measurement of 
chromatic dispersion. 

\subsection{Results: single-cladding waveguides and multi-cladding waveguides with small refractive index contrast}
The simulation sections showed that either an increase of $h_c$ or an increase of the $t_1$ and/or $t_2$ in a waveguide with a small 
refractive index contrast produces a vertical up-shift of the dispersion profile. This up-shift in turn results in a change of the 
zero dispersion location. Figure 19 shows the measured chromatic dispersion for the quasi-TE mode in waveguides where, through an 
increase either of the core height or of the cladding layer thickness, the location of the $\lambda_0$s is progressively shifted. The 
blue line shows the measured dispersion in a single-cladding waveguide with $w_c = 1.5 $ $\mu $m and $h_c = 0.725 $ $\mu $m - there 
are two $\lambda_0$s located at $\lambda_{01}$ $\approx$ 1150 and $\lambda_{02}$ $\approx$ 1720 nm. The dispersion of a small 
refractive index contrast waveguide with $w_c = 1.5 $ $\mu $m, $h_c = 0.75 $ $\mu $m, and $t_1 = t_2 = 0.1 $ $\mu $m is plotted with a 
black line. Because of adding the layers $t_1 $ and $ t_2$, the dispersion profile is vertically up-shifted: the $\lambda_{02}$ is 
shifted to 1980 nm, while the $\lambda_{01}$ lies outside of our measurement spectral range but, by extrapolation, is expected 
to be at $\sim$1120 nm. The red line shows the dispersion of a small refractive index contrast waveguide with $w_c = 1.3 $ $\mu $m, 
$h_c = 0.75 $ $\mu $m, and $t_1 = t_2 = 0.2 $ $\mu$m where the profile is again vertically up-shifted: the $\lambda_{02}$ is at 
$\sim$2150 nm. Finally, for a waveguide with $w_c = 1.3 $ $\mu $m, $h_c = 0.75 $ $\mu $m, and $t_1 = t_2 = 0.4 $ $\mu $m the 
profile in green shows that there is only one zero dispersion wavelength ($\lambda_{01}$ $\approx$ 1480 nm) at the spectral region 
of our measurements. 

\begin{figure}[htbp]
\centerline{\includegraphics[width=1.0\columnwidth]{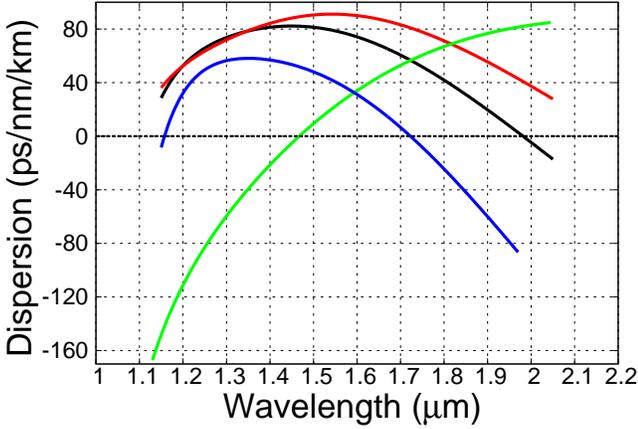}}
\caption{Dispersion for the quasi-TE mode for a single-cladding waveguide (blue) and multi-cladding waveguides with small refractive index contrast 
(black, red and green).}
\label{fig:Figure20}
\end{figure}

The waveguide parameters for the cases illustrated in black and green lines are identical with the ones in dotted red and solid black 
lines in the simulation of Figure 11, respectively. The qualitative agreement 
between both results is excellent: for example, $\lambda_{01} \sim$ 1480 nm in the green spectrum in Fig. 19 and 1550 nm in the 
corresponding simulation in Fig. 11 (black solid line). Furthermore, $\lambda_0$s $\sim$ are located at 1120 and at 1980 nm in the black spectrum 
in Fig. 19 and at 1280 and at 2060 nm in the corresponding simulation in Fig. 11 (red dashed line). Even though the shape agrees well 
(but measured profiles are slightly blue-shifted in comparison with simulation), quantitatively, the measured dispersion is higher 
(80 vs. 45 ps/nm/km) than the simulated one. We believe that this is primarily due to the refractive index uncertainty and also to 
the uncertainty of the exact waveguide dimensions (estimated to have an error of 5 - 10 nm).

\subsection{Results: Multi-cladding waveguides with large refractive index contrast}

As demonstrated in the simulation sections, a large refractive index contrast is required to have a more arbitrary control of the 
chromatic dispersion, in particular flat profiles centered at the telecommunication window. Figure 20 shows the measured chromatic 
dispersion of the quasi-TE mode of multi-cladding waveguides with a refractive index arrangement similar to Fig. 1(e). The inset shows 
the sketch of the transversal section of the fabricated waveguides. The profile 
in blue is obtained with waveguide dimensions $h_c = 0.75 $ $\mu $m, $w_c = 1.8 $ $\mu $m, $t_1 = 60 $ nm and $t_2 = 120 $ nm: Even 
though the dispersion profile has comparable flatness ($\pm$ 11 ps/nm/km variation between 1300 and 1800 nm) as in the multi-cladding 
waveguide with small refractive index contrast, the advantage of having large contrast is that the average dispersion is lower and 
the mode confinement is tighter. The profile in black ($h_c = 0.75 $ $\mu $m, $w_c = 1.8 $ $\mu $m, $t_1 = 90 $ nm and $t_2 = 120 $ nm)
has a larger $t_1$, so that similar flatness to the one in blue is produced while the average dispersion is lower. Finally, a strong 
increase of $t_2$ and a decrease of $t_1$ ($h_c = 0.75 $ $\mu $m, $w_c = 1.8 $ $\mu $m, $t_1 = 40 $ nm and $t_2 = 200 $ nm) results 
in a much better dispersion optimization - the profile in green is very flat (31 $\pm$ 3.2 ps/nm/km) in the region between 1300 and 
1800 nm. This flatness is comparable to the ones in state-of-the-art photonic crystal fibers \cite{26,28}.

\begin{figure}[htbp]
\centerline{\includegraphics[width=1.0\columnwidth]{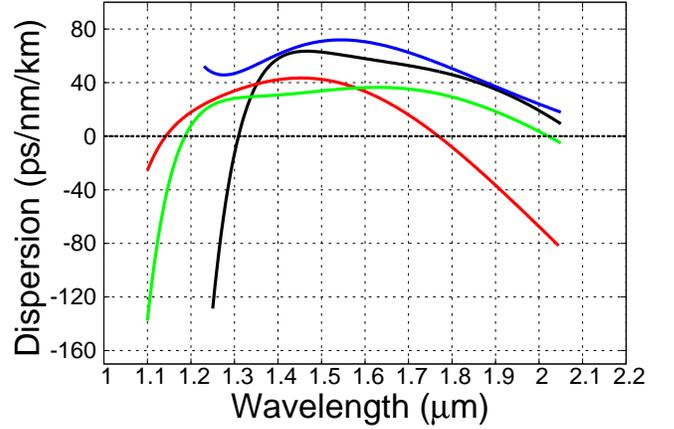}}
\caption{Dispersion for the quasi-TE mode in multi-cladding waveguides with large refractive index contrast.}
\label{fig:Figure21}
\end{figure}

Figure 21 illustrates that even though the dimensions of the fabricated $Si_X N_Y$ waveguides were chosen to produce flat dispersion profiles
for the quasi-TE propagation mode, the multi-cladding arrangement is also advantageous to flatten dispersion for quasi-TM propagation. This is 
well exemplified with the profile depicted in red which is obtained with the waveguide dimensions $h_c = 0.75 $ $\mu $m, 
$w_c = 1.3 $ $\mu $m, $t_1 = 40 $ nm and $t_2 = 200 $ nm and has a dispersion variation of $\pm$ 2 ps/nm/km between 1210 and 1400 nm. 
The precision of the whole engineering process (fabrication and dispersion characterization) is illustrated by changing by a small amount 
the waveguide dimensions. The profile in black is obtained with identical dimensions than the one in red except that $w_c = 1.5 $ $\mu $m: 
it has similar flatness and shape in comparison with the profile in red, as expected from simulations. Furthermore, from simulations it 
is expected that a decrease of $t_2$ produces a vertical up-shift of the dispersion profile. In this context, a waveguide having dimensions 
$h_c = 0.75 $ $\mu $m, $w_c = 1.5 $ $\mu $m, $t_1 = 60 $ nm and $t_2 = 120 $ nm produces the profile in blue which is vertically up-shifted 
by 40 ps/nm/km from the previous cases. Finally, the profile in green is with a waveguide with identical dimensions than the one in blue 
except that $w_c = 1.3 $ $\mu $m: both profiles have very similar shapes as expected from simulations. 

\begin{figure}[htbp]
\centerline{\includegraphics[width=1.0\columnwidth]{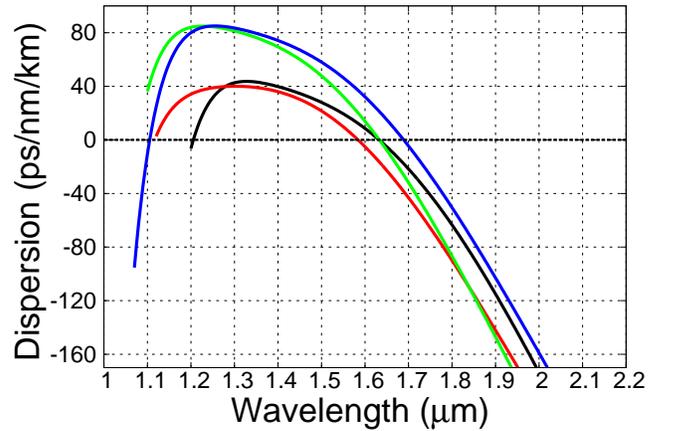}}
\caption{Chromatic dispersion for the quasi-TM mode in a multi-cladding dispersion-engineered waveguides.}
\label{fig:Figure22}
\end{figure}

\subsection{Experimental results: supercontinuum generation in multi-cladding silicon nitride waveguides}
 
Multi-cladding waveguides with dispersion engineered characteristics should reduce the power threshold for nonlinear interactions. To 
verify this we propagate a train of 85 fs pulses with a central wavelength at 1560 nm with the aim of generating supercontinuum (SC). 
Pulses with 90 pJ energy are coupled into the quasi-TE mode of the 1.4 cm long $Si_X N_Y$ waveguide with a dispersion flattened profile: 
$h_c = 0.75 $ $\mu $m, $w_c = 1.8 $ $\mu $m, $t_1 = 40 $ nm and $t_2 = 200 $ nm (its dispersion is depicted with a green line in Figure 20). 
Two OSAs (one operating from 350 to 1200 nm while the other from 1200 to 2400 nm) are used to acquire the SC spectra. The generated SC 
is fairly flat and extends nearly over three octaves (from 700 to 2400 nm, limited by the operating bandwidth of the OSA). At 800 nm a 
strong dispersive wave is generated. This is the broadest SC generated in an integrated waveguide \cite{40}. The dependency of the bandwidth 
with the dispersion profile is illustrated in the light-blue spectrum which was obtained for the quasi-TM mode of a 1.4 cm long waveguide 
with $h_c = 0.815 $ $\mu $m, $w_c = 1.5 $ $\mu $m, and $t_1 = t_2 = 0 $ (its dispersion is depicted in Figure 25) \cite{41}. Due to the 
much larger and less flat dispersion, the SC spans only between 1250 and 2300 nm. To verify that the different SC bandwidths come from 
the different dispersion characteristics, the propagation losses are measured by substracting the transmission power in waveguides with 
different lengths. The losses (peaking at 1520 due to H-N bonds) are similar for both types of multi-cladding waveguides as can be seen 
in the green line for the waveguide with flat and low dispersion and the light green line for the one with large dispersion.

\begin{figure}[htbp]
\centerline{\includegraphics[width=1.0\columnwidth]{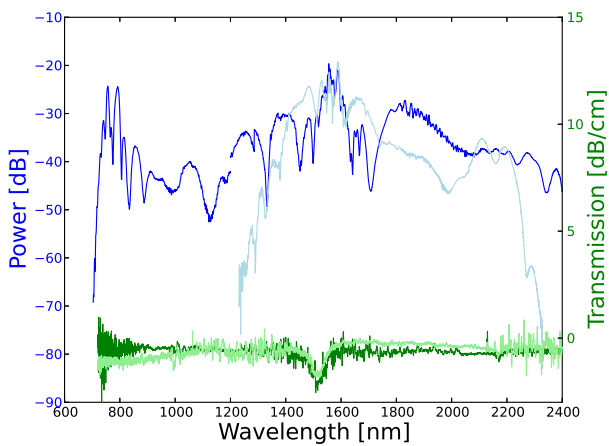}}
\caption{SC spectrum generated in a multi-cladding waveguide with flat and low dispersion (blue line) and with large dispersion (light blue line)).}
\label{fig:Figure23}
\end{figure}

\section{Conclusion}

Dispersion engineering by geometrical and refractive index optimization in silicon nitride waveguides was numerically and experimentally 
investigated. We demonstrated that by incorporating cladding layers with a refractive index ratio of $n_2/n_1 \sim 1.4$, very flat 
dispersion profiles can be engineered. For this refractive index ratio, random fluctuations of the waveguide parameters preclude producing
ultra-flat dispersion profiles, although profiles with $\pm$ 0.5 ps/nm/km variation are still numerically demonstrated over a 740 nm 
bandwidth. Single- and multi-cladding waveguides were fabricated and their dispersion profiles measured, showing that flattened profiles 
are accessible for both quasi-TE and quasi-TM mode propagation. The profiles can be modified with high predictability by changing the core height or 
the thickness of the cladding layers. A flat dispersion profile with a variation of $\pm$ 3.2 ps/nm/km variation over a 500 nm 
bandwidth is obtained in a multi-cladding waveguide fabricated with a refractive index contrast of 1.37. The dispersion control characteristics 
demonstrated here are particularly useful for improving the performance of a number of nonlinear devices and a specific case is 
analyzed by demonstrating almost three octaves SC generation. 
This should open new avenues for the advancement of linear and nonlinear functionalities on a chip. 

\section{Acknowledgment}
We would like to acknowledge the financial support from the Federal Ministry of Education and Research (BMBF) under grants 03Z2AN11 
and 03Z2AN12. Beate Kerpen is gratefully acknowledged for useful comments on the manuscript.

\section{Appendix}

To determine the chromatic dispersion, we divide the interference pattern in sub-sections containing an appropriate number of maxima (around 2-4) 
and perform a sinusoidal fit with the following function:

\begin{equation}
A + B\times\sin(e \times f + g).
\end{equation}

It is assumed that in each sub-section $\phi$ is well represented by $ e \times f + g$, where $e = \frac{d \phi}{d f}$, $f$ is the frequency 
and $g$ is a phase for that subsection. The coefficient $B$ is the amplitude of the sinusoidal function, while $A$ is an offset to fit the spectrum. 
The advantage of this fitting function is that it involves the whole interference spectrum, and not only the region where the maxima are 
located \cite{33}. The black dots in Figures 23(a), 24(a), and 25(a) show examples of retrieved $\frac{d \phi}{d f}$ values over a bandwidth of $\sim 120$ 
THz. The data is then fitted with a third-order polynomial shown with a red line. By taking the derivative of this polynomial the chromatic 
dispersion (shown in solid black lines in Figures 23(b), 24(b), and 25(b)) is finally retrieved. The case in Figure 23 corresponds to a waveguide 
with small refractive-index contrast (as in Fig. 1c) and having $h_c = 0.75 $ $\mu $m, $t_1 = t_2 = 0.4 $ $\mu $m. Figure 24 illustrates the very 
flat dispersion profile case shown in Figure 20 ($h_c = 0.75 $ $\mu $m, $w_c = 1.8 $ $\mu $m, $t_1 = 60 $nm, and $t_2 = 0.2 $ $\mu $m). Finally, the 
Figure 25 shows the profile for the quasi TM mode of a single cladding waveguide with $h_c = 0.815 $ $\mu $m and $w_c = 1.5 $ $\mu $m ($t_1 = t_2 = 0 $).
Note that in all the cases, the measured $\frac{d \phi}{d f}$ shows a dip centered at around 200 THz (1520 nm) which comes from transmission losses 
due to H-N bonds (see transmission spectrum in Fig. 22). While in the flattened case, $\frac{d \phi}{d f}$ changes merely from 4 
to 6.5 ps$^{-1}$ (indicating that the dispersion is low and constant), in the other two cases the change is much bigger (from -6 to -12.5 ps$^{-1}$ and 
from from -6.5 to -19.5 ps$^{-1}$) accounting for a larger chromatic dispersion. Figure 24(a) shows a region between 210 and 220 THz where the measured 
data cannot be fitted by the third order polynomial - this might be due to the extreme flatness in this waveguide. Furthermore, Figures 23(a), 24(a), 
and 25(a) indicate that the measurements are expected to be more accurate in determining the zero dispersion wavelength (that occurs when the 
derivative of $\frac{d \phi}{d f}$ is zero) than in determining the maximum of dispersion (that occurs when the second derivative of $\frac{d \phi}{d f}$ 
is zero).

\begin{figure}[htbp]
\centerline{\includegraphics[width=1.0\columnwidth]{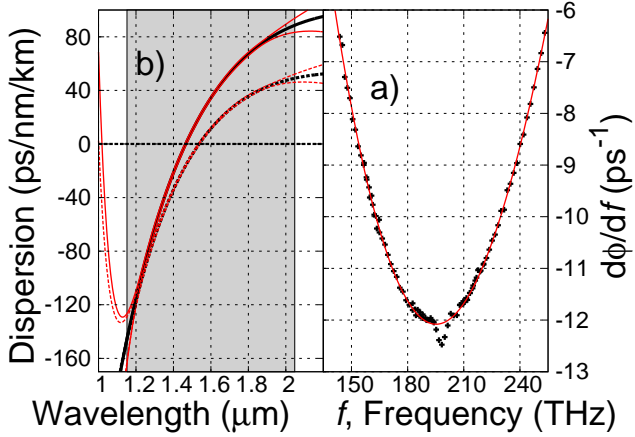}}
\caption{(a) Black dots: measured $d \phi /d f$; red line: fitting of $d \phi /d f$ using a third order polynomial. (b) Dispersion in a multi-cladding waveguide with small refractive index contrast. Shadow area: region covered by the measurement.}
\label{fig:Figure24}
\end{figure}

\begin{figure}[htbp]
\centerline{\includegraphics[width=1.0\columnwidth]{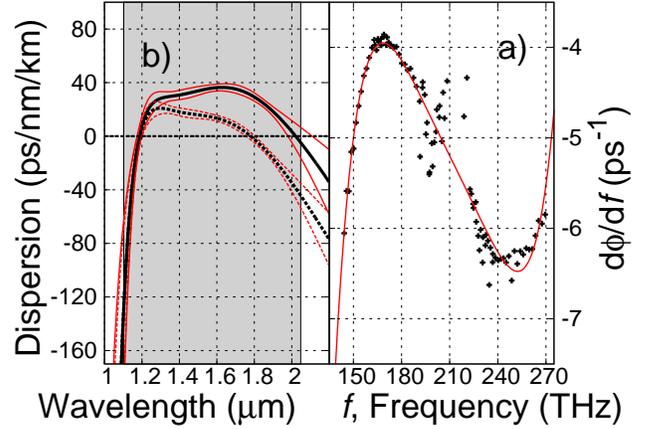}}
\caption{(a) Black dots: measured $d \phi /d f$; red line: third order polynomial fitting. (b) Chromatic dispersion in a multi-cladding waveguide with large refractive index contrast. Shadow area: region covered by the measurement.}
\label{fig:Figure25}
\end{figure}

\begin{figure}[htbp]
\centerline{\includegraphics[width=1.0\columnwidth]{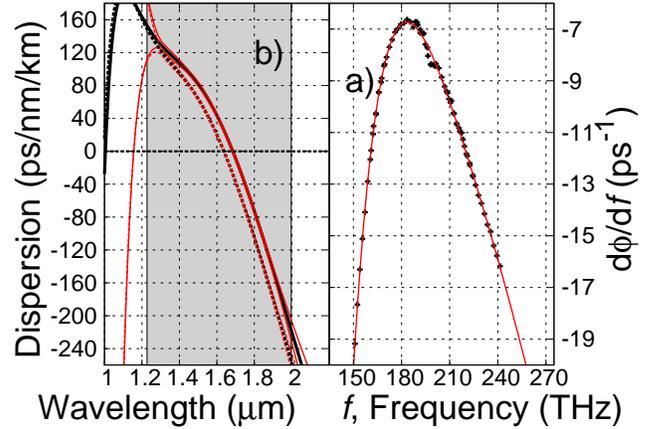}}
\caption{(a) Black dots: measured $d \phi /d f$; red line: third order polynomial fitting. (b) Chromatic dispersion in a single-cladding waveguide. Shadow area: region covered by the measurement.}
\label{fig:Figure26}
\end{figure}

When making the polynomial fitting of $\frac{d \phi}{d f}$, we calculated the curves where there is 95\% probability that the measured
$\frac{d \phi}{d f}$ values will lie within this confidence interval. From these curves, we calculated the red curves in Figures 23(b), 24(b), and 25(b) - the 
resulting dispersion within the 95\% confidence interval is nearly indistinguishable from the original curves (except at the edges) for the large 
dispersion cases and there is a small difference for the flat and low dispersion case. For comparison purposes, the 
dotted black lines in Figs. 23(b), 24(b), and 25(b) show the measured dispersion before doing the substraction of the group delay obtained when 
the waveguide is removed. The difference between solid and dotted lines illustrates the contribution of the group delay from uncompensated optical components
in the interferometer, which is important only at long wavelengths. 

A comparison of the chromatic dispersion when it was measured over large bandwidths 
($\sim 1000$ nm) and over much shorter ones (e.g. 300 nm) shows that covering both $\lambda_{0} $s is the optimal condition for a more accurate dispersion
measurement.


\end{document}